\documentclass[11pt,twoside]{article}
\usepackage{asp2014}

\aspSuppressVolSlug
\resetcounters

\begin{document}

\title{Modeling \hfill software \hfill solutions \hfill and \hfill computation \hfill facilities \hfill for \hfill FAIR \hfill access}

% full name: Sara Bertocco
\author{Sara~Bertocco}
\affil{INAF~-~Osservatorio Astronomico di Trieste, 34143 Trieste, Italy; \email{sara.bertocco@inaf.it}}\
\paperauthor{Sara~Bertocco}{sara.bertocco@inaf.it}{0000-0003-2386-623X}{INAF}{OATs}{Trieste}{}{34131}{Italy}

% leave these next few aindex lines commented for the editors to enable them. Use Aindex.py to generate them for yourself.
% first presenting author should be the first entry for bold-facing the author index page-reference
%\aindex{Bertocco,~S.}
% remove/add as you need

% leave the ssindex lines commented for the editors to enable them, use Index.py to suggest yours
%\ssindex{FOOBAR!conference!ADASS 2022}
%\ssindex{FOOBAR!organisations!ASP}

% leave the ooindex lines commented for the editors to enable them, use ascl.py to suggest yours
%\ooindex{FOOBAR, ascl:1101.010}
  
\begin{abstract}
We are in the era of the Big Data. In Astronomy and Astrophysics, 
the massive amounts of data generated are, as of today, in the Peta-scale 
if not already in the Exa-scale. In the near future, we will see the data 
collected size and complexity grow further constantly, 
setting new challenges for data processing, reduction and analysis. This will 
pose new needs in terms of software and hardware solutions but also in 
terms of new models for resource management, access and sharing. 
In Astronomy and Astrophysics, in the environment 
of the International Virtual Observatory Alliance (IVOA), a big work has already 
been done with regards to data, gaining 
complete data FAIRness. In this paper, a model is proposed, 
based on the IVOA architecture, 
for software and hardware solutions for data analysis. The goal of this 
model is to build a cloud to access Astronomy and Astrophysics resources 
following the FAIR principles.
\end{abstract}

\section{Introduction}
The Astro community\footnote{Astronomy, Astrophysics, Astroparticle and 
Astroinformatics} is in the era of the Big Data. The TeraByte scale was reached in the 1990s
with large digital sky surveys and the multi-PetaByte scale 
was reached by the late 2010s, as documented in \citet{2013pss2.book..223D}. 
To manage such a large amount of data, the Virtual Observatory (VO) was developed.
At the beginning, it was the vision that astronomical 
datasets and other resources should work as a seamless whole. 
Now it is a global data cloud providing access in a transparent way to geographycally distributed
data archives from both ground-based observatories and space-based missions,
thanks to the implementation of standards and services developed in the IVOA
framework. IVOA is 
based on an architecture matching the FAIR (Findable, Accessible, Interoperable, 
Reusable) principles.

Now we are moving or maybe we are already in the Exa-scale era and in the near future, 
we will see the amount of data grow further. Big projects like the 
High-Luminosity Large Hadron Collider (HL-LHC) and the Square Kilometre Array (SKA) 
are each expected to produce, in the next decade, up to an exabyte of data per year. 
The VO is a management tool for the geographically distributed big data access and new 
storage architectures and solutions
are under evaluation and development such as in \citet{2020EPJWC.24504019B and 2021EPJWC.25102060D}. 
On the other hand, the amount and rates 
of generated data and their complexity, pose new challenges in terms of data processing, 
reduction and analysis as well as the need of hardware solutions overcoming the traditional ones. 
New models are needed to manage these software and hardware solutions in a geografically 
distributed infrastructure, matching data with suitable analysis software solutions and 
adequate hardware where to run. 

This paper starts with a brief recap of which are the FAIR guiding principles in 
Section \ref{fair_princ}, then the VO is intrduced in Section \ref{VO_arch_sec}, 
focusing on its FAIR architecture, 
composed of a resource layer, a registry, access protocols, a client layer and a set 
of standards providing the semantic needed to gain interoperability. In Section \ref{the_model},
is proposed a model for software and hardware facilities, similarly of what 
already done for the data in the IVOA framework. 
In the following sections, are reviewed existing building blocks 
of this model, used so far without any standardization.
In Section \ref{repos}, repositories as software resource layer; 
in Section \ref{registries}, the Astrophysics Source Code Library 
(ASCL)\footnote{https://ascl.net/}, an example of software registry. 
In Section \ref{metadata}, is introduced the software metadata schema question, 
highlighting the CodeMeta project, moving toward a standardization.
In Section \ref{facilities_model}, are outlined existing ways of modeling 
computing facilities, mainly in conjunction with ``run software'' use cases, with
some details on the IVOA Execution Planner.
In Section \ref{use_case}, it is suggested a use case of the model proposed in Section \ref{the_model} and 
in Section \ref{escape}, it is the description of the work done in the EU project ESCAPE matching, 
in a specific pattern,
the proposed use case. In Section \ref{the_conclusion}, the reasoning done in this article is summarized
and a futuristic use case is proposed for the reproducibility of scientific results
and their publication.

\section{The FAIR Guiding Principles}
\label{fair_princ}
With FAIR we mean Findable, Accessible, Interoperable and Reusable. In ~\citet{2016NatSD...360018W} 
are formally published for the first time ``The FAIR Guiding Principles for scientific data management 
and stewardship'', a concise and measureable set of principles to be used as guideline to improve 
data re-usability. These principles put specific emphasis on {\bf enhancing the ability of machines to 
automatically find and use the data}, in addition to supporting its reuse by individuals. They are: \\ 
\underline{To be Findable}: {\bf F1.} (meta)data are assigned a {\bf globally unique and persistent identifier};
{\bf F2.} data are described with {\bf rich metadata} (defined by R1 below); {\bf F3.} 
metadata clearly and explicitly include the identifier of the data it describes; {\bf F4.} 
(meta)data are {\bf registered or indexed} in a searchable resource. \\
\underline{To be Accessible}: {\bf A1.} (meta)data are retrievable by their identifier using a {\bf standardized 
communications protocol}; {\bf A1.1} the protocol is open, free, and universally implementable; {\bf A1.2} the protocol allows for an authentication and authorization procedure, where necessary; {\bf A2.} metadata are accessible, even when the data are no longer available.\\
\underline{To be Interoperable}: {\bf I1.} (meta)data use a formal, accessible, shared, and broadly 
applicable {\bf language for knowledge representation}; {\bf I2.} (meta)data use {\bf vocabularies} 
that follow FAIR principles; {\bf I3.} (meta)data include qualified references to other (meta)data. \\
\underline{To be Reusable}: {\bf R1.} meta(data) are richly described with a plurality of 
accurate and relevant attributes; {\bf R1.1} (meta)data are released with a clear and accessible 
data usage license; {\bf R1.2} (meta)data are associated with detailed provenance; {\bf R1.3} 
(meta)data meet domain-relevant community standards.

\section{Virtual Observatory and IVOA FAIR Architecture}
\label{VO_arch_sec}
The Virtual Observatory (VO) was born as the vision that astronomical
datasets and other resources should work as a seamless whole. Now it has evolved
in a global data cloud
providing access, in a transparent way, to data archives geographycally distributed, from both 
ground-based observatories and space-based missions.

The International Virtual Observatory Alliance (IVOA) is 
a global collaboration of separately funded projects 
within which a set of standards are debated, defined and promoted to 
provide data centers with a standard framework
for publishing and delivering services using their data.
The implementation of these standards for data, their metadata
and exchange methods, plus a registry, which lists available services
and what can be done with them\footnote{\url{https://ivoa.net}} fosters the VO
creation, based on an architecture matching the FAIR principles, Figure \ref{VO_arch_fig}.

\articlefigure[width = 333pt]{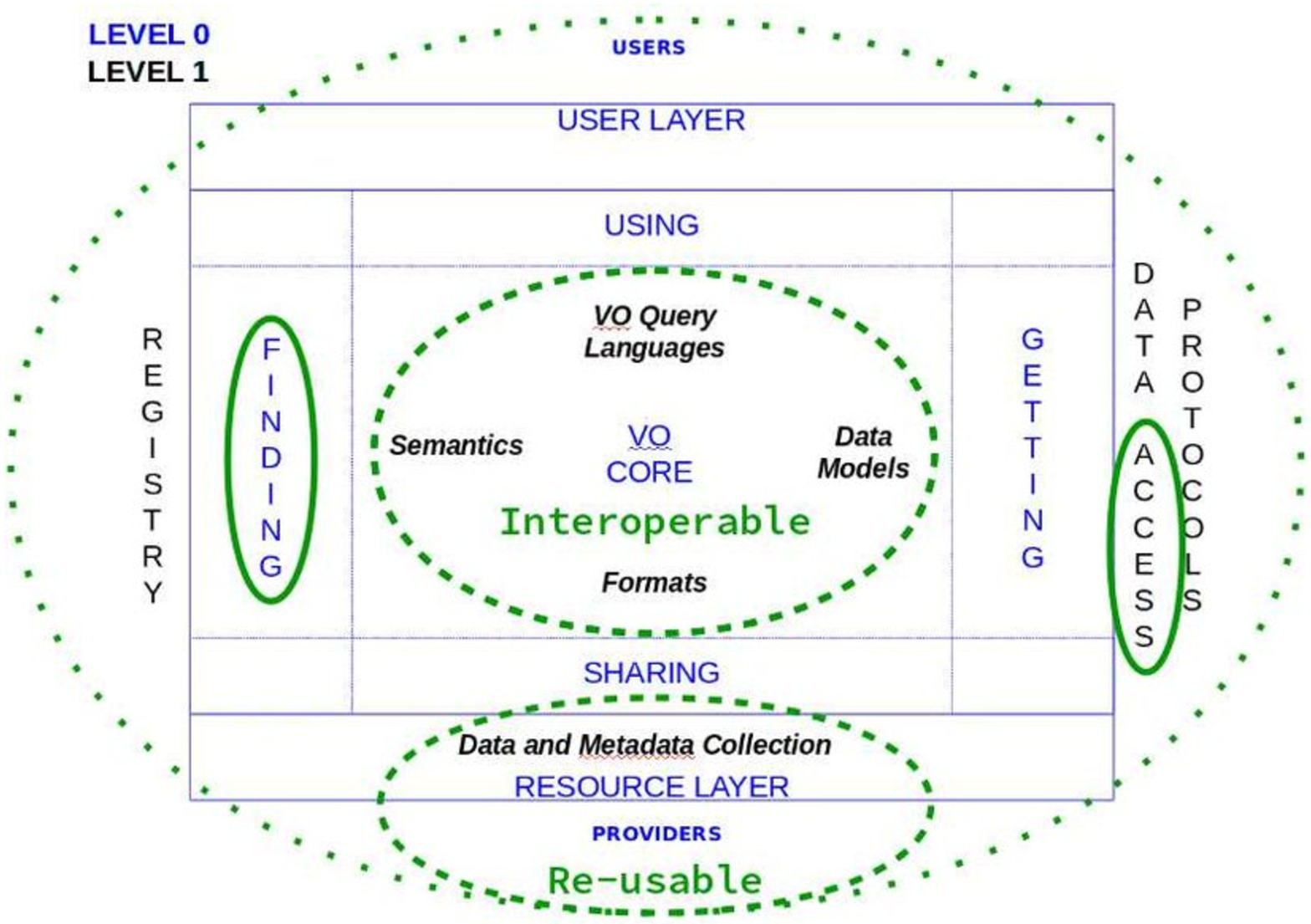}{VO_arch_fig}{The FAIR concepts superimposed (as ellipses)
to conceptual schema (Levels 0 and 1) of the VO architecture. 
This picture was used in several technical meetings since late 2017, and first published in \citet{2020ASPC..522..707P}.}

As explained in \citet{2021ivoa.spec.1101D}, interoperability fits the core of the Level 1 
IVOA Architecture, while the pillars of Registry and Data Access Protocols embody the 
Findability and Accessibility. Reusability pops out of the full architecture, taking 
advantage of the entire ecosystem the VO architecture defines. See also \citet{2021arXiv211114468M}.
It has been shown that the FAIR Guiding 
Principles are totally fulfilled by any Astro archive compliant with the IVOA standards, 
matching all requirements summarized in section \ref{fair_princ}. See \citet{2022arXiv220310710O}, \citet{2020ASPC..522..707P}.
Moreover, archives compliant with the IVOA standards can be federated in the VO framework 
leading to the result wanted since the 
beginning: give to the user tools to access and analyze astronomical data 
spread all over the world, in different data centers, transparently.

\section{FAIR model for software and computing facilities}
\label{the_model}
Starting from the IVOA architecture, we can think of modeling in the same way software resources and hardware facilities with:
a) a Resource layer, b) a Registry, c) use of Unique Identifiers, d) Standard metadata description of the resources,
e) Standars for access, query, exchange resources, f) a Language for knowledge representation.
If we analyze our daily work, a good part of these components 
is already present and widely used. Developers use repositories to store software and
manage development works;
researchers archive data and products or register reports or publications
in a variety of web services. All these works are done using
common mechanisms, such as metadata descriptions of saved things (data, software,
reports, etc.) but there are not standard definitions, nor seamless ways to use services, nor interoperability.

\section{Software resource layer: Repositories} 
\label{repos}
Software repositories are widely used by software developers and users. 

{\bf https://github.com}: is an Internet hosting service for software 
development and
version control using git. It provides mainly development tools like distributed version control,
access control, bug tracking, software feature requests, task
management, continuous integration, and wikis for every project.

{\bf https://zenodo.org}: is an open repository where researchers can deposit papers, data
sets, software, reports, etc. Zenodo records are described with metadata compliant 
with DataCite's Metadata Schema minimum and recommended terms, with a few additional 
enrichements. Zenodo provides also a globally unique and persistent identifier, a DOI
(Digital Object Identifier) minted for each 
submission\footnote{https://about.zenodo.org/principles/}

{\bf https://hub.docker.com} Docker Hub is a library and community for storing and
sharing container images. Through the Docker Hub ``Explore'' page in Figure \ref{dockerhub}  
it is provided the ability to filter a search operation based on the hardware architecture
suitable to run the software. 
\articlefigure[width = 333pt]{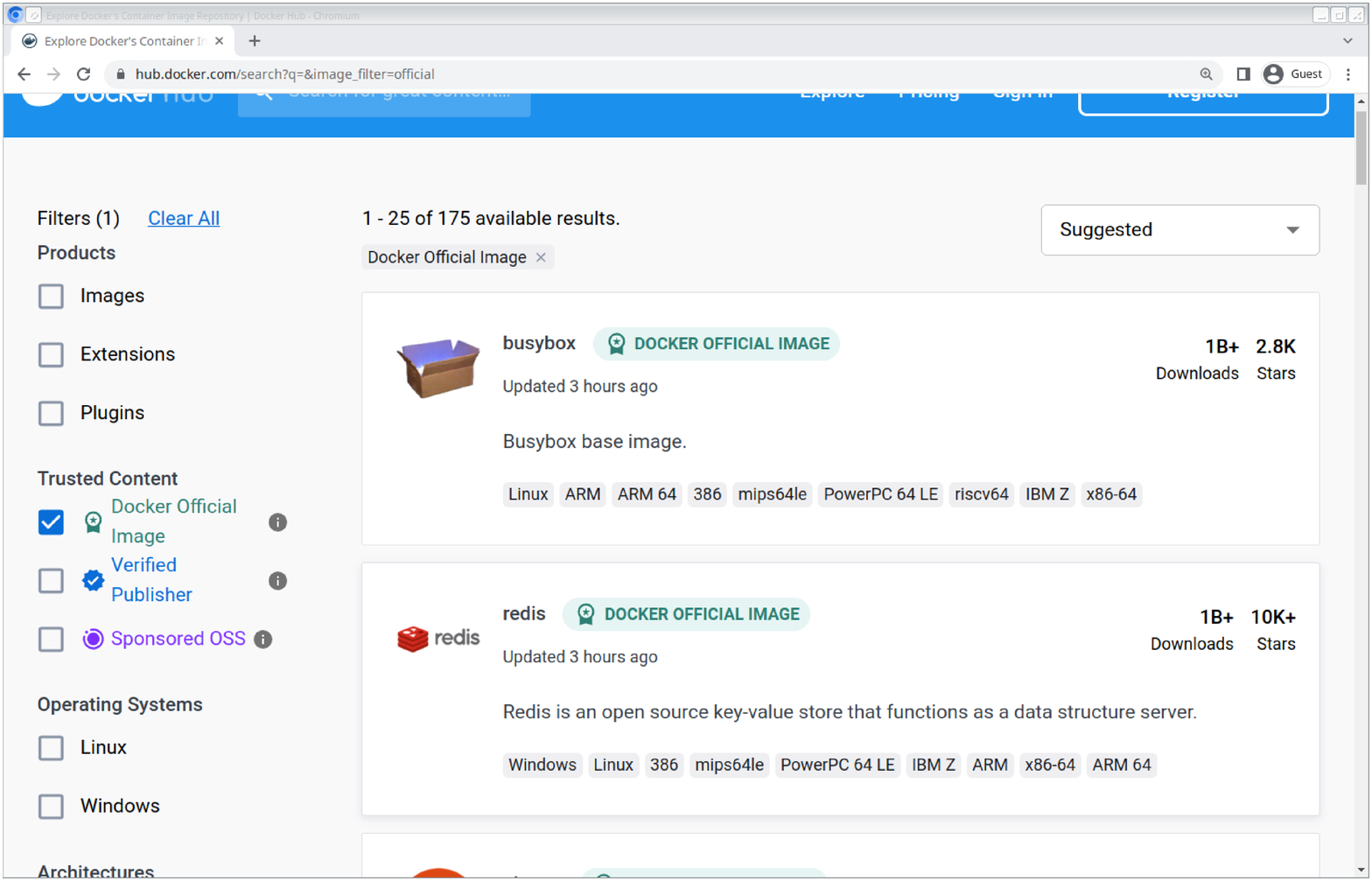}{dockerhub}{The\ ``Explore'' page of the Docker Hub
repository.}
This means that the software metadata description is rich enough
to retain information about the ability to run specific software.

\section{A software registry: ASCL}
\label{registries}
The ``Astrophisics source code library'' (ASCL) is a free online registry for source code of 
interest for astronomers and astrophysicists, \citet{2020ASPC..522..731A}. 
It is citable through a custom unique identifier but it mints also DOIs and is indexed 
by ADS\footnote{The SAO/NASA ADS
(Astrophysics Data System) is an online database of million astronomy and physics 
papers from both peer reviewed and non-peer reviewed sources.}.
ASCL improves metadata code description to make better discoverability and proper crediting 
supporting both codemeta.json and CITATION.cff formats, \citep{2022ASPC..532..627T}.

{\bf codemeta.json} is a 
minimal metadata schema for science software and
code, in JSON and XML formats\footnote{\url{https://codemeta.github.io/}. \url{https://github.com/codemeta/codemeta}}. 

{\bf CITATION.cff} Citation File Format (CFF) files
are plain text files with human- and machine-
readable citation information for software (and
datasets)\footnote{https://citation-file-format.github.io/}. 

Code developers can include these files in
their repositories to let others know how to
correctly cite their software, \citep{T01_adassxxxii}. % For Editors: Citation of tutorial "Using the Astrophysics Source Code Library: Find, cite, download, parse, study, and submit" A.Allen - Please verufy the citation correctness

\section{Software metadata description and opportunity of standardization}
\label{metadata}
Software metadata description is a challenging task. Software has to be 
considered from different points of view: credits, i.e. citations,
versions and dependencies (for running purpose), discoverability (related to the 
ability to run a specific task) and so on.
Different use cases need a different description. For this reason, different software repositories, 
and scientific domains denote the information about code in different ways.
They have software description,
archival and in some cases citability, but in a non standard way,
nullifying 
the ability of machines to
automatically find and use software depending from its use case.
Recalling the IVOA architecture
and remembering the key principials to gain FAIRness, software resource FAIRness can be
obtained defining and using unique identifiers, rich metadata with domain-relevant community standards, language
for knowledge representation and communication protocols.

The {\bf CodeMeta Project} is moving toward a software metadata description standardization. 
It started by comparing the software metadata used across multiple repositories,
creating a ``CodeMeta Metadata Crosswalk'' (a map between the metadata fields 
used by different software repositories, registries and archives). The crosswalk 
was then used to generate a set of software metadata with the aim 
to standardize the exchange of software metadata across repositories and organizations. 

\section{How to model computing facilities?}
\label{facilities_model}
A repository of resource facilities has no sense. What can be usefully suggested is a 
registry of resource facilities where computing resource providers can register their resources.
This requires 
a metadata description of the computing facilities. 
At present, to the author's knowledge, no metadata descriptions of computing facilities or 
registries of available resources are in use, but at most rankings of the best performing 
HPC clusters in the world. 
However there are examples of tools for service discovery and usage configuration for computing resources 
belonging to a grid, \citet{2017JPhCS.898i2042A}.

In the IVOA framework, thanks to the support of the European
Project ESCAPE, see section \ref{escape}, a document note\footnote{An IVOA note 
is a document expressing suggestions from and opinions of the authors. It is intended to share 
best practices, possible approaches, or other perspectives on interoperability with the Virtual 
Observatory and should not be referenced or otherwise interpreted as a standard specification.},
called Execution Planner, has been written
about the possibility to match software with suitable computing facilities.

\subsection{The IVOA Execution Planner}
\label{exec_planner}
The IVOA Execution Planner (IVOA-EP) interface is a HTTP web-service interface, 
that provides a simple way to discover and access computing services with the 
purpose to run specific software\footnote{\url{https://github.com/ivoa/ExecutionPlannerNote}, Retrieved: November 25, 2022}. 
It is designed as a REST\footnote{REST (REpresentational State Transfer) is a software architecture that describes an  
interface for resource access through HTTP standard methods.} interface, based on two classes of objects: 
{\bf computing tasks} and {\bf computing services}, to be able to answer the 
question: ``Which computing service can I use to run this task?''. 
This question can be formally expressed as an HTTP request GET to an 
ExecutionPlanner service providing a tasktype parameter in the URL, as in the following example.
\begin{verbatim}
HTTP GET /accepts?tasktype=uri://docker-container
\end{verbatim}  
The question can also be more complicated: ``Can I run this Docker container here?'' 
In this case, the GET request should contain, in the payload, ``task specifc content'' in a
``taskinfo element'', that carrys the ``fully qualifed name'' of the container image, as in the following example:
\begin{verbatim}
{"tasktype": "uri://docker-container",
 "taskinfo": {
      "image": "docker.io/example:1.0"}
}
\end{verbatim}
The service will answer yes or no. If yes, assuming the container image meets some 
acceptance criteria, the IVOA-EP service would reply including 
a ``servicetye parameter'' to indicate the type of service and a ``serviceinfo parameter'' 
to indicate the endpoint URL to access the service. Example:
\begin{verbatim}
{"reponseword": "YES",
 "servicetype": "uri://docker-uws",
 "serviceinfo": {
      "endpoint": "http://example.org/docker-uws-runner",}
}
\end{verbatim}

\section{A use case}
\label{use_case}  
Given a metadata description of a software, or more generally, a computing task description, 
and having also a computing facility/service characterization (both through metadata), 
a match can be realized (algorithmic in a complex solution or simply a coupling 
task-``service facility'' in a simpler one) to provide a mechanism to run software on a suitable
hardware. In figure \ref{use_case_fig} a
diagram of the functionality that can be obtained.
\articlefigure[width = 325pt]{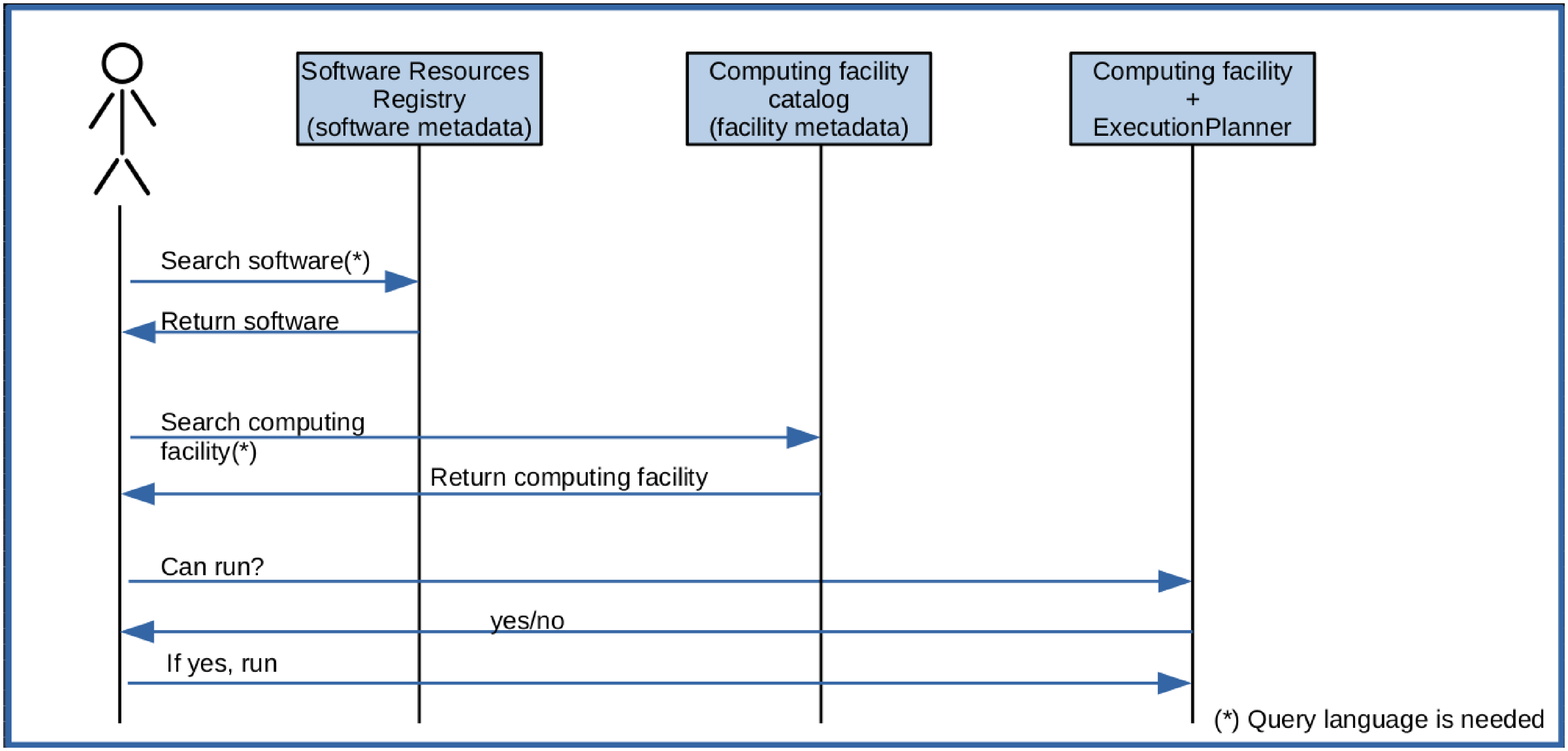}{use_case_fig}{Use case: a user searches for software in a repository, then,
having a software description, that in the simplest case is a unique identifier, searches 
in a registry a suitable facility where run the software. At the end he is able to run the software.}

\section{The work done in ESCAPE}
\label{escape}
The ESCAPE EU project performed a basic implementation of the use case in section \ref{use_case}, using
tools that could be basic building blocks of the model proposed in this paper.

ESCAPE \footnote{ESCAPE: European Science Cluster of Astronomy \& Particle physics ESFRI research infrastructures. \url{https://www.projectescape.eu} and \url{https://cordis.europa.eu/project/id/824064},\\  Retrieved: 25 November 2022} 
brings together astronomy and particle physics
communities to address fundamental challenges in data-driven research, inspired by the
goals and needs of major European research infrastructures, or ESFRIs. \citep{swinbank_john_d_2022_6044637} Aims of the project are:
\begin{itemize}
\vspace{-0.2cm}\item to support the implementation of the European Open Science Cloud by fostering {\bf open data management according to FAIR principles}
\vspace{-0.3cm}\item  to develop joint multiwavelength and multimessenger data processing and analysis capabilities
\vspace{-0.3cm}\item  to enable {\bf interoperability} between facilities
\end{itemize}
\vspace{-0.2cm}These goals are addressed developing a range of services, including data infrastructure, a software repository, Virtual Observatory integration, and a citizen science programme. 

\subsection{OSSR, the ESCAPE Open-source Scientific Software and Service Repository}
\label{ossr}
One of the actions of the ESCAPE project is to provide tools necessary to share science products, 
promote open science and maximize cross-fertilization by software re-use and co-development 
through its Open-source Scientific Software and Service Repository (OSSR).

The ESCAPE repository is implemented as a curated community in Zenodo web service, 
labelled ``escape2020'' \citep{2022arXiv221200499V}. A Zenodo curated community 
is a topic centric collection where the community curator (generally the creator) 
has the ability to decide what uploaded records are allowed to be associated to the 
community. This way, a sort of review process can be implemented. 
Moreover, the community collection is automatically exported 
via OAI-PMH\footnote{The Open Archives Initiative Protocol for Metadata Harvesting (OAI-PMH) 
is a low-barrier mechanism for repository interoperability. Data Providers are repositories 
that expose structured metadata via OAI-PMH. Service Providers then make OAI-PMH service 
requests to harvest that metadata. Reference: \url{https://www.openarchives.org/pmh/},  
Retrieved: November 25, 2022} making it citable and discoverable. 

One of the aims of the OSSR is to collect software responding to the ESFRI's use cases. 
To support the ESFRI's software developers and users, several tools to enable a
complete software life-cycle have been integrated.
In the OSSR, an API has been developed to programmatically access it, retrieve records 
and publish content. Functions are provided to map and crosswalk metadata between the CodeMeta schema 
adopted for the OSSR and Zenodo internal schema \citep{vuillaume_thomas_2022_6826881}.

\subsection{OSSR metadata}
\label{ossr_metadata}
All the OSSR records are described through the Zenodo metadata, but, for
uploads containing software, it is strongly recommended to add to the record a
metadata file following the CodeMeta Project schema, specifically designed to describe 
scientific software and code. The inclusion of the codemeta.json file
in the submitted record and its validity are crucial to ensure FAIR compliance.
The OSSR implementation provides a Codemeta 
generator\footnote{\url{https://escape2020.pages.in2p3.fr/wp3/codemeta-generator/}, \newline
Retrieved: 25 November 2022} and a Codemeta Validator 
\footnote{\href{https://escape2020.pages.in2p3.fr/wp3/eossr/docstring_sources/eossr_cli/eossr-metadata-validator.html}
{\nolinkurl{https://escape2020.pages.in2p3.fr/wp3/eossr/docstring_sources/eossr_cli/} \hfill \nolinkurl{eossr-metadata-validator.html}}, Retrieved: 25 November 2022}. 

\subsection{ESCAPE ESAP: the ESFRI's Science Analysis Platform}

ESAP, the ESFRI's Science Analysis Platform, aims to provide a gateway to the full range of ESCAPE services and functionalities, 
while being customizable and adaptable to the needs of particular science goals or 
research infrastructures \citep{swinbank_john_d_2022_6044637}. 
It is conceived as a modular and extensible architecture that makes it more a science platform 
toolkit than a monolithic science platform. 
Currently a prototype is available providing, 
among other features, a basic implementation of the use case in Section \ref{use_case},
underlying the model of hardware and resource management proposed in this paper, also if lacking
the standards.
It offers the ability to:
\begin{itemize}
\vspace{-0.2cm}\item query and select datasets through VO tools and the implementation of 
VO standards ADQL (\citet{2008ivoa.spec.1030O}) to manage archive queries 
and SAMP (\citet{2015A&C....11...81T}) to collect data through external tools.
\vspace{-0.3cm}\item browse the software stored in the OSSR through the provided APIs (see Section \ref{ossr}).
\vspace{-0.3cm}\item run the selected software in a suitable computing infrastructure.
\end{itemize}

\section{Conclusions and future view}
\label{the_conclusion}
This paper proposes to model software and hardware resources by implementing the same
architecture used in the IVOA to manage data. Pillars are: a) software 
repositories and software metadata description; b) hardware facilities 
metadata description; c) a registry for both; d) unique identifiers; e) a 
language for knowledge representation; f) information exchange protocols.
For the software
metadata description standardization, the work is someway already started 
(Section \ref{metadata}, CodeMeta project). For the hardware facilities the road seems to be longer,
but the effort is started (Section \ref{exec_planner}). A use case is already implemented, 
in a basic shape, in the EU ESCAPE project (Sections \ref{use_case},\ref{escape}).
For the future, this model could be used for scientific result reproducibility, matching
data, analysis software and hardware resources where run the analysis.
A starting point for a really nice application is in
\citet{2020ASPC..527..225W}. Given the implementation of the model and the use case proposed, 
the scientific result could be published in digital papers providing some result
``live'' reproducibility. This would be of great help to referees of software-related 
publications who have difficulties in validating, testing, reproducing the presented 
results as discussed in \citet{B02_adassxxxii}.

\acknowledgements  ESCAPE, the European Science Cluster of Astronomy \& Particle Physics ESFRI 
Research Infrastructures, has received funding from the European Union's Horizon 2020 research and innovation programme under Grant Agreement no. 824064.

\bibliography{I01}

% if we have space left, we might add a conference photograph here. Leave commented for now.
% \bookpartphoto[width=1.0\textwidth]{foobar.eps}{FooBar Photo (Photo: Any Photographer)}

\end{document}